# Transcranial Bipolar Direct Current Stimulation
# of the Frontoparietal Cortex Reduces Ketamine-Induced Oscillopathies:
# A Pilot Study in the Sedated Rat

Caroline Lahogue[1,2,3] and Didier Pinault[1,2,3,*].

(1) INSERM U1114, Neuropsychologie cognitive et physiopathologie de la schizophrénie, Strasbourg, France.
(2) Université de Strasbourg, Strasbourg, France.
(3) Fédération de Médecine Translationnelle de Strasbourg, FMTS, Faculté de médecine, Strasbourg, France.

**Running title:**

Frontoparietal anodal tDCS reduces ketamine-induced oscillopathies

(∗) **Corresponding author:**

Didier Pinault, pinault@unistra.fr

**ABSTRACT**

During the prodromal phase of schizophrenia with its complex and insidious clinical picture, electroencephalographic recordings detect widespread oscillation disturbances (or oscillopathies). Neural oscillations are electro-biomarkers of the connectivity state within systems. A single systemic administration of ketamine, a non-competitive NMDA glutamate receptor antagonist, transiently reproduces the oscillopathies with a clinical picture reminiscent of the psychosis prodrome. This acute pharmacological model may help the research and development of innovative treatments against the psychotic transition. Transcranial electrical stimulation is recognized as an appropriate non-invasive therapeutic modality since it can increase cognitive performance and modulate neural oscillations with little or no side effects. Therefore, our objective was to set up, in the sedated adult rat, a stimulation method able to normalize the ketamine-induced oscillopathies. Unilateral transcranial frontoparietal anodal stimulation by direct current (<+1 mA) was applied in ketamine-treated rats. A concomitant electroencephalographic recording of the parietal cortex measured the stimulation effects on its spontaneously-occurring oscillations. A 5-min bipolar anodal tDCS immediately and quickly reduced, significantly with an intensity-effect relationship, the ketamine-induced oscillopathies at least in the bilateral parietal cortex. A duration effect was also recorded. These preliminary neurophysiological findings are promising for developing a therapeutic proof-of-concept against neuropsychiatric disorders.

## INTRODUCTION

Effective treatments against chronic schizophrenia without side effects are still missing[1, 2]. Its development takes years with the occurrence of prodromal symptoms associated with attention-related sensorimotor and cognitive deficits[3], dysfunctional brain networks[4, 5], and widespread oscillation disturbances[6, 7]. These prodrome-related oscillopathies include an excessive amplification of broadband gamma-frequency (30-80 Hz) oscillations[8] and a reduction in the density of sleep slow-wave oscillations and spindles[9-14]. Neural oscillations, naturally implicated in attentional and integrative processes, are biomarkers of the connectivity state within systems. Proton magnetic resonance spectroscopy reveals, during the prodrome, a decrease in the glutamate and glutamine levels[15, 16]. They are correlated with gray matter volume in the frontoparietal (FP) system. These findings support the glutamate hypothesis of schizophrenia[2, 17].

Aberrant amplification of broadband gamma oscillations can be reproduced in cortical and subcortical structures in healthy humans and rodents after a single systemic administration, at a psychotomimetic dose, of the N-methyl-D-aspartate glutamate receptor antagonist ketamine[18-21]. The state and function of cortical and subcortical networks are altered, including in the FP corticothalamic system, which plays an essential role in attentional and integrative processes. Furthermore, the ketamine-elicited gamma hyperactivity decreases the ability of cortico-thalamo-cortical networks to integrate incoming information[22, 23]. In pentobarbital-sedated rats, ketamine transiently reduces the power of slow-wave oscillations and spindles by switching the firing pattern of both thalamic relay and reticular neurons from the burst mode to the single action potential mode[24]. Furthermore, clozapine, one of the most effective antipsychotic medications currently available, especially in treatment resistant patients with schizophrenia[25, 26], prevents the ketamine effects on thalamocortical slow-wave oscillations and spindles[24]. Therefore, the ketamine-induced oscillopathies represent translational electrical biomarkers for cerebral network disorders with prognostic and therapeutic potential, a hope for the research and development of innovative treatments against the psychotic transition.

Transcranial direct current stimulation (tDCS) is recognized as an appropriate non-invasive therapeutic modality since it can increase cognitive performance and modulate neural oscillations with little or no side effect[27-30]. This stimulation makes it possible to modulate the physiological or pathological cortical and subcortical activity in humans[31-33]. In healthy volunteers at rest, functional magnetic resonance imaging reveals that bipolar FP or fronto-temporal anodal tDCS modulates corticostriatal and corticothalamic connectivity[32, 33]. In addition, at the level of the primary motor cortex, tDCS can modulate the different nodes of the cortico-thalamo-cortical circuit. In addition, FP tDCS makes it possible, in patients with schizophrenia that are resistant to antipsychotics, to reduce positive[34] and negative[35] symptoms. As well, in patients with schizophrenia, anodal tDCS can reduce the gamma event-related synchronization[36]. Also, the use of tDCS has a great potential in the treatment of cognitive symptomatology in early psychosis[37]. Nevertheless, the tDCS-induced immediate and downstream effects are transient and the parameters of tDCS remain to be further investigated in order to better understand their impact on network and cellular activities.

Our objective was to set up a preclinical, experimental bipolar FP tDCS design allowing the refinement of the parameters under well-controlled conditions while recording the ongoing parietal EEG oscillations. The bipolar format (nearby stimulating electrodes) was privileged as it results in more localized current flow than the monopolar DCS (two remote stimulating (brain) and reference (e.g, body) electrodes), which stimulates a larger amount of brain tissue[38]. For that, we used the pentobarbital-sedated rat, a model of slow-wave sleep with spindle-like activities and bouts of gamma oscillations[24], which made it possible to quantitatively apprehend the stimulation effects on the ketamine-induced oscillopathies. The present, conceptually- and data-driven pilot study shows that a unilateral FP anodal tDCS was substantially efficient to reduce them.

## METHODS AND MATERIALS

### Animals and drugs

Fourteen Wistar adult male rats (285-370 g) were used with procedures performed under the approval of the Ministère de l'Education Nationale, de l'Enseignement Supérieur et de la Recherche. Ketamine was provided from Merial (Lyon, France).

### Surgery under general anesthesia

Deep general anesthesia was initiated with an intraperitoneal injection of pentobarbital (60 mg/kg). An additional dose (10-15 mg/kg) was administered when necessary. Analgesia was achieved with a subcutaneous injection of fentanyl (10 μg/kg) every 30 minutes. The anesthesia depth was continuously monitored using an electrocardiogram, watching the rhythm and breathing, and measuring the withdrawal reflex. The rectal temperature was maintained at 36.5 °C (peroperative and protective hypothermia) using a thermoregulated pad. The trachea was cannulated and connected to a ventilator (50% air–50% O2, 60 breaths/min). Under local anesthesia (lidocaine), an incision of the skin of the skull was done and the periosteum was removed to set the skullcap bared and to perform the stereotaxic positioning of the stimulating and recording electrodes on the FP skull. The general anesthesia lasted about 2 hours, the time needed to complete all the surgical procedures.

### Analgesic pentobarbital-induced sedation

At the end of the surgery, the body temperature was set to and maintained at 37.5°C. The analgesic pentobarbital-induced sedation was initiated about 2 h after the induction of the deep anesthesia and maintained by a continuous intravenous infusion of the following regimen (average quantity given per kg and per hour): Pentobarbital (4.2 ± 0.1 mg), fentanyl (2.4 ± 0.2 μg), and glucose (48.7 ± 1.2 mg). In order to help maintain stable the ventilation and to block muscle tone and tremors, a neuromuscular blocking agent was used (d-tubocurarine chloride: 0.64 ± 0.04 mg/kg/h). Local wound anesthesia is maintained. The cortical EEG and heart rate were under continuous monitoring to adjust, when necessary, the infusion rate to maintain the sedation. The EEG recordings began 2 hours after the beginning of the infusion of the sedative regimen.

### Unilateral frontoparietal tDCS combined with bilateral cortical EEG

For the unilateral, bipolar frontoparietal tDCS, we used pellet Ag/AgCl electrodes (Warner Instruments), 1.5 mm in diameter and 3 mm in height. The electrodes were positioned on the left side of the skull, the cathode above the frontal area (relative to the bregma: anterior: 5 mm; lateral: 1 mm) and the anode above the parietal area (anterior: 3 mm; lateral 3 mm) (Fig. 1A1, A2). In an attempt to reduce the inhomogeneities of electrical conductivity of the skin-skull interface[39], the skull was slightly drilled at the electrode placement areas (~2 mm in diameter), a strategy that secured the electrode position. The electrodes were positioned on wet sponges (NaCl, 0.9%) 2 mm in diameter and 1.0 mm in thickness (Fig. 1A2, A3). Drops of saline solution were regularly applied on the sponges in order to keep them moist, a strategy to minimize the electrical shunting effects, which was expected to maintain as stable as possible the current flow during the application of the stimulating current. The electric current was supplied using a Master-8 stimulator (A.M.P.I) equipped with an isolation unit to deliver a constant current. At the end of the experiment, the animals were killed during a lethal injection of Euthasol (pentobarbital, 400 mg/kg).

For the bilateral cortical EEG, four recording Teflon-sheathed silver wires (diameter: 200 $\mu$m) were implanted in the skull. After a slight drill hole, the recording section of the wires was in contact with the internal plate of the skull. The two, right and left, active electrodes were placed in the parietal skull over the primary somatosensory cortex (from bregma: 2.3 mm posterior; 5 mm lateral), and the references (ground mode) were positioned on the occipital ridges. The EEG signals (0.1-800 Hz) were acquired using an ultralow-noise differential amplifier (AI 402, x50; Molecular Devices). All signals were sampled at 10 kHz 16-bit (Digidata 1440A with pCLAMP10 Software, Molecular Devices).

## Data analysis

Analysis software packages Clampfit v10 (Molecular Devices) and SciWorks v10 (Datawave Technologies) were used. Spectral analysis of baseline EEG oscillations was performed with the fast Fourier transformation (FFT, 0.5 Hz resolution). The power of EEG activities was analyzed in 3 frequency bands: delta-(1–4 Hz), sigma-(10–17 Hz, spindles), gamma-(30–80 Hz) frequency oscillations. For each band, the total power was the sum of all FFT values. Statistical analyzes were performed using the software R. Parametric tests were used to assess the significance of the results: paired student t test, one-way analysis of variance with a Tukey's post-hoc test HSD ("honestly significant difference", significance level p <0.05). Each animal was its own control.

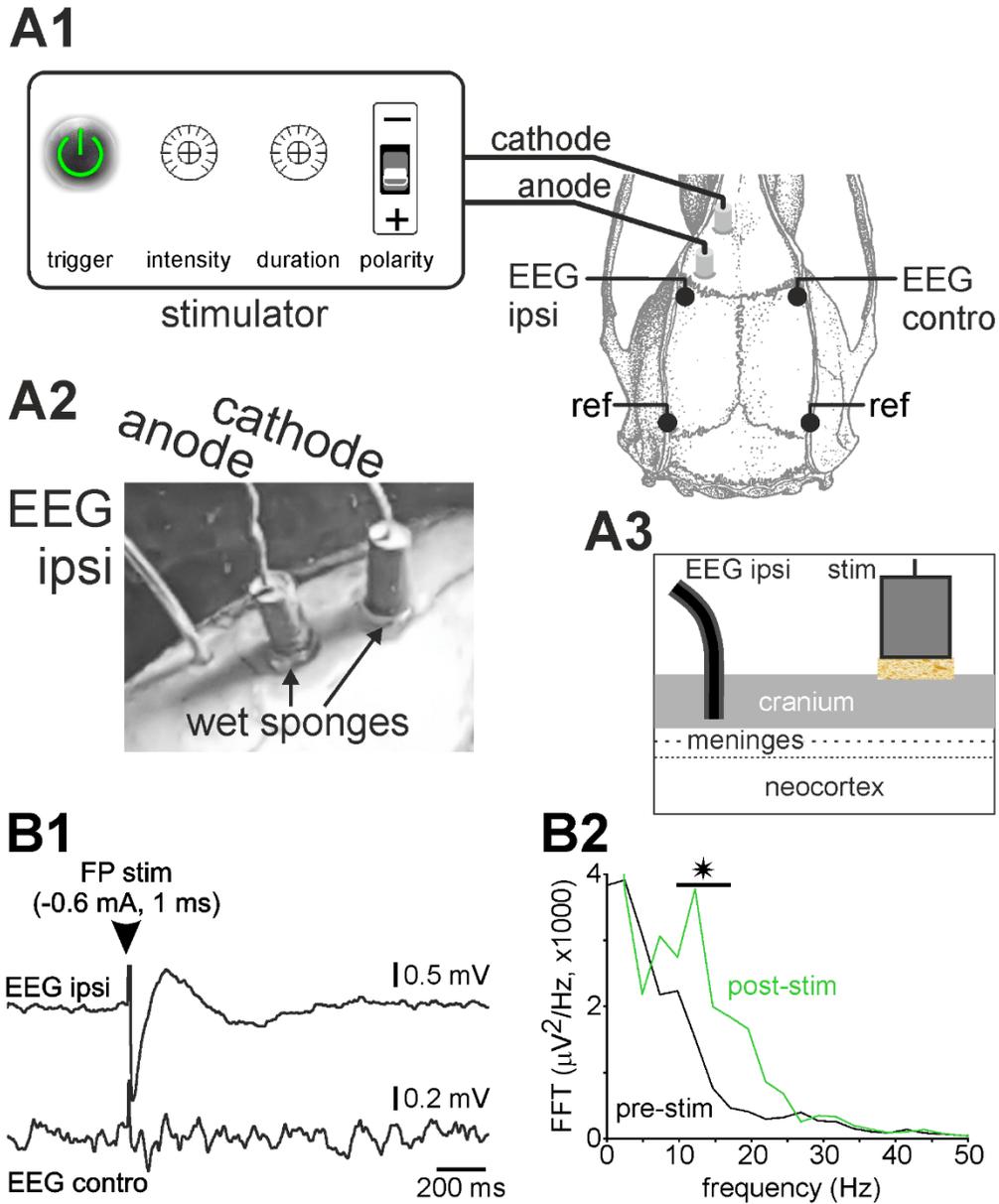

**Figure 1: Experimental design. (A1)** Dorsal view of the rat skull showing the location of the EEG recording electrodes positioned on the parietal somatosensory cortex, and the stimulation electrodes, the cathode at the frontal level and the anode at the parietal level. The references of the EEG electrodes are positioned on the occipital ridges. **(A2)** Photography of the bipolar stimulation electrodes and of the ipsilateral EEG electrode (EEG ipsi) relative to the location of the stimulation electrodes. **(A3)** Diagram illustrating the location, non-invasive from the point of view of the brain, of the EEG electrodes, the surface of each electrode in contact with the inner plate of the skull, and the stimulation electrodes, each lying on a sponge soaked in saline solution and pressed on the surface of the skull. **(B1)** Evoked bilateral parietal cortex EEG response (averaged 10 sweeps) following a bipolar frontoparietal electrical stimulation (-0.6 mA, 1 ms) in a pentobarbital sedated rat. The stimulation immediately (~1 ms) evokes a typical evoked potential at the ipsilateral EEG electrode (EEG ipsi), and a sigma-frequency oscillation lasting ~1 second at the contralateral EEG electrode (EEG contro). **(B2)** Spectral analysis of the contralateral EEG before (pre-stim) and after (post-stim) the frontoparietal stimulation. Student's test (*, $p < 0.05$).

## RESULTS

Under our experimental ketamine-free (control) conditions, the EEG recordings displayed spontaneous and predominant oscillations in the delta frequency band (1-4 Hz, or slow waves) accompanied by oscillations in the sigma band (10-17 Hz, or "spindle-like" activities)[24, 40]. These oscillatory activities had characteristics qualitatively similar to the slow waves sleep with spindles recorded in free-behaving rats in stage II sleep. The slow-wave sleep-type oscillations could be interspersed with smaller and faster oscillations including, among others, broadband gamma- and higher-frequency oscillations.

| Parameters | N | Protocol |
|---|---|---|
| Limit intensity | 2 | from +/-0.2 mA to +/-1 mA, 1 min |
| Polarity | 1 | -1 mA, 1 min |
|  | 2 | +1 mA, 1 min |
| Duration | 1 | Control 0mA |
|  | 1 | 0 mA, 1 min |
|  | 1 | +0.5 mA, 1 min |
|  | 1 | +0.5 mA, 2 min |
|  | 1 | +0.5 mA, 5 min |
| Timing | 1 | +0.5 mA, 5 min, 10 min after keta injection |
|  | 1 | +0.5mA, 5 min, 5 min after keta injection |
|  | 2 | +0.5 mA, 5min, 8 min after keta injection |
| Intensity | 3 | 0 mA, 5min, 8 min after keta injection |
|  | 3 | +0.25 mA, 5 min, 8 min after keta injection |
|  | 3 | +0.5 mA, 5 min, 8 min after keta injection |

**S1: Summary table of the parameters tested.** The list of the tested parameters is, from top to bottom, in the chronological order. N, number of rats for each condition.

The parametrization of the bipolar FP tDCS began according to an empirical and pragmatic approach (S1). The experimental conditions were stable and reliable enough giving the possibility to adjust the stimulation parameters during the course of every experiment and from one to the next experiment, a refinement strategy that gives potentially useful results with a reasonable number of animals. More specifically, each experiment was followed by a spectral analysis of strategic EEG segments and a debriefing to decide the stimulation parameters to apply on the next rat. Four rats were used to optimize the efficiency of the wet sponge to minimize the potential electrical shunting effect of the electrode-skull interface (Fig. 1A1-A3) to get a stable and reliable stimulation effect. This was assessed on the basis of the ipsi- and controlateral potential responses evoked following a 1-ms pulse of electrical microstimulation with an intensity varying from -0.03 to -0.60 mA (2 rats, Fig. 1B1,B2). After a short latency (~1ms), an evoked potential was recorded in the ipsilateral EEG and a sigma-frequency oscillation lasting ~1 s on the controlateral parietal cortex (Fig. 1B1). The evoked controlateral oscillation, highlighted after averaging, was visible from -0.30 mA. The spectral analysis revealed a significant increase (p<0.05) in the power of the sigma oscillations, which corresponded to an evoked spindle-like activity (Fig. 1B2).

Our objective being to set up a bipolar FP tDCS capable of reducing or normalizing the effects of ketamine on spontaneously-occurring neuronal oscillations, all the results presented in the following were obtained in the ketamine condition, that is, after a single subcutaneous administration of ketamine at a subanesthetic and psychotomimetic dose (2.5 mg/kg)[19]. It is worth reminding that, in the sedated rat, ketamine fleetingly decreases the power of spindles (sigma-frequency oscillations) and increases that of broadband gamma oscillations with a peak effect 15-20 minutes after its systemic administration[24]. A partial or total recovery is usually observed 60-80 minutes later. In the present study, an overview of the overall effect of ketamine is visible in the Figure 2. Among the 14 rats, 4 were excluded from the data analyzes because the sponge interface pads were not wet enough and/or properly positioned, which impacted the quality of the tDCS current flow and caused many artifacts.

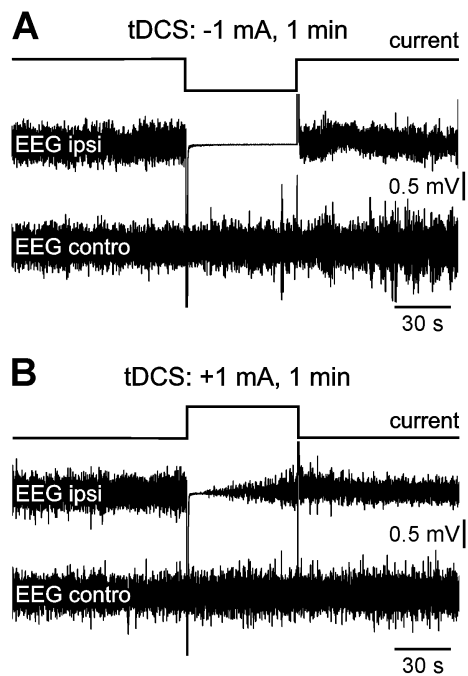

**S2: Saturation of the ipsilateral EEG amplifiers by a 1-min frontoparietal cathodal or anodal stimulation with a relatively high intensity.** The polarity effect of the tDCS is tested on two pentobarbital sedated rats **(A and B)** 20 minutes after a subcutaneous administration of ketamine (2.5 mg/kg) for the cathodal stimulation (-1 mA, A) and 10 minutes after for the anodal stimulation (+1 mA, B). The amplifier saturation was observed only in the ipsilateral EEG.

## Bipolar frontoparietal anodal tDCS

We started to seek an intensity capable of modulating immediately, quickly and substantially the pattern of EEG oscillations. For this, a 1-min FP tDCS with an increasing intensity was applied either from 0 to +1 mA, or from 0 to -1 mA (with 0.2 mA increments) on a first rat, and from +0.5 to +1 mA (0.5 mA increment) on a second rat. In each rat, the first stimulation was applied 10 min after the administration of ketamine, that is, ~5 minutes before its peak effect (at 15-20 min postinjection) on cortical gamma oscillations. The minimum time interval in between two successive tDCS was 8 min. A tDCS of +1 or -1 mA was, immediately, able to fully or transiently saturate the amplifier of the ipsilateral EEG. The EEG took on the appearance of an isoelectric trace, whereas the controlateral EEG was slightly or not affected (S2). The spectral analysis of the cortical EEG activities reveals (Fig. 2): i) a transient normalization in the ketamine-induced gamma hyperactivity for ~5 minutes, an effect that was more remarkable in the ipsilateral than the controlateral EEG; and ii) a decrease in the power of the concomitant spindle-like activity lasting approximately 2 min followed by a transient increase for ~5 min, an effect also more remarkable in the ipsilateral than the controlateral EEG. These results were reproducible in another rat, leading us to further investigate the bipolar FP anodal tDCS with an intensity inferior to +1mA and by increasing the duration (>1 min).

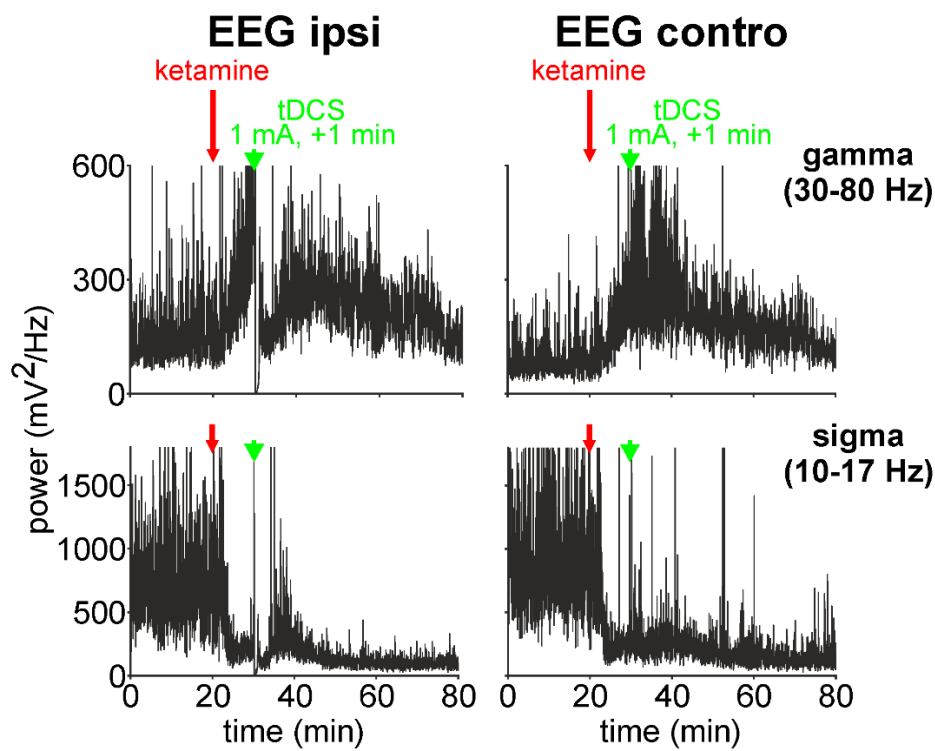

**Figure 2: Anodal tDCS of +1mA for 1 minute transiently modulates the power of the ketamine-induced bilateral, gamma hyperactivity and sigma hypoactivity.** The stimulation is applied 10 minutes after the systemic administration of ketamine (2.5 mg/kg at time 20 min), that is, when the ketamine-induced sigma hypoactivity and gamma hyperactivity are well installed.

### Duration effect

Therefore, the duration effect was assessed with a FP tDCS of +0.5 mA in 2 rats, one rat per duration (2 and 5 min). The stimulation was applied 10 minutes after the ketamine administration. The spectral analysis was carried out for 2 minutes 5 minutes after the end of the tDCS, that is, at the peak effect time of the ketamine-elicited gamma hyperactivity. The results are presented in the figure 3. The effectiveness of the tDCS in reducing the ketamine-induced gamma hyperactivity increased when increasing the duration. More specifically, a full normalization was recorded in both the ipsi- and the controlateral EEGs after an anodal tDCS (+0.5 mA) lasting 5 min.

Regarding the spindle-like activities, the tDCS +0.5 mA, with a duration of 2 or 5 min, tended to normalize the ketamine-induced reduction in spindle power in the bilateral EEG (Fig. 3). The effectiveness of the stimulation in the ipsilateral EEG was higher with a duration of 5 min than with 2 min, whereas both durations had almost the same effect in the controlateral cortex. A complete normalization (non-significant relative to the saline condition) was recorded in the ipsilateral EEG with a tDCS of 5 min. So, these observations led us to set the duration of the tDCS at 5 minutes.

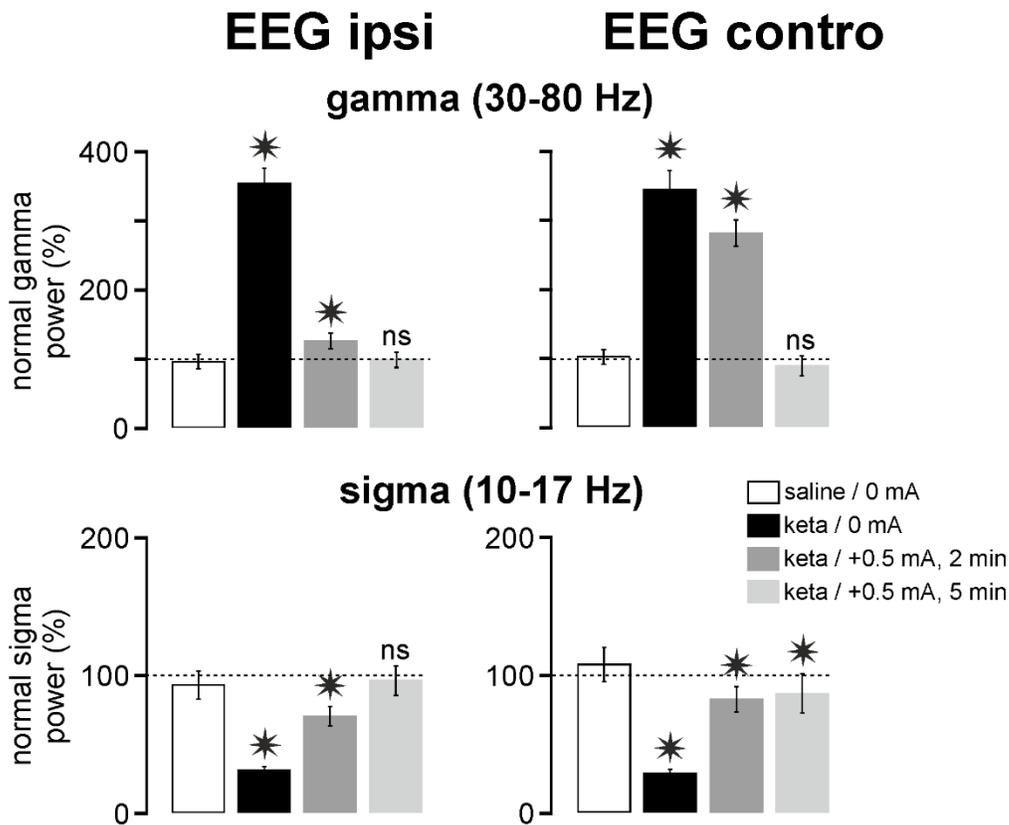

**Figure 3: Duration effect of tDCS on ketamine-induced gamma hyperactivity and sigma hypoactivity.** The spectral analysis (fast Fourier transform) is performed for 2 minutes 5 minutes after the end of the tDCS. Data are normalized. The ketamine effect (keta/tDCS 0 mA) and each of the keta/tDCS conditions +0.5 mA (2 or 5 min) are compared to the "saline/0 mA" control (1 rat per condition, 60 values/rat, each rat being its own control). Student's test (ns, non-significant; *, p <0.05).

## When to apply the bipolar tDCS?

So far, we applied our stimulation 10 min after the systemic administration of ketamine, that is, when its peak effect on cortical gamma oscillations started to emerge. In a previous study, it was demonstrated that clozapine, one of the most effective antipsychotic medications currently available, prevented the ketamine effects[24]. So, with the idea to prevent the ketamine-induced oscillopathies, we wanted to test, in another rat, whether an earlier tDCS was capable in deleting the ketamine peak effect. For that, we applied the 5-min tDCS +0.5 mA 5 min after the ketamine administration (Fig. 4). At the time of ketamine injection, the EEG was relatively synchronous, containing delta oscillations and spindles. The bipolar anodal tDCS was applied when (5 min postinjection) the ketamine effects started to be visible (Fig. 4). During the stimulation, atypical slow waves occurred accompanied with many artifacts (amplifier saturations) especially in the ipsilateral EEG. The atypical waves and artifacts disappeared immediately after the end of the stimulation. The tDCS significantly suppressed the ketamine-induced peak of gamma hyperactivity (p < 0.001, at 35-40 min) completely in the ipsilateral EEG and partially in the controlateral EEG (Fig. 5A,B). In this rat, the stimulation was ipsilaterally so powerful that it significantly reduced the gamma power below the normal value (100%) recorded under the saline condition. The same anodal tDCS +0.5 mA tardily and significantly (p < 0.001 at 85-90min) reduced the ketamine-induced spindle hypoactivity in the bilateral cortical EEG without reaching a normalization.

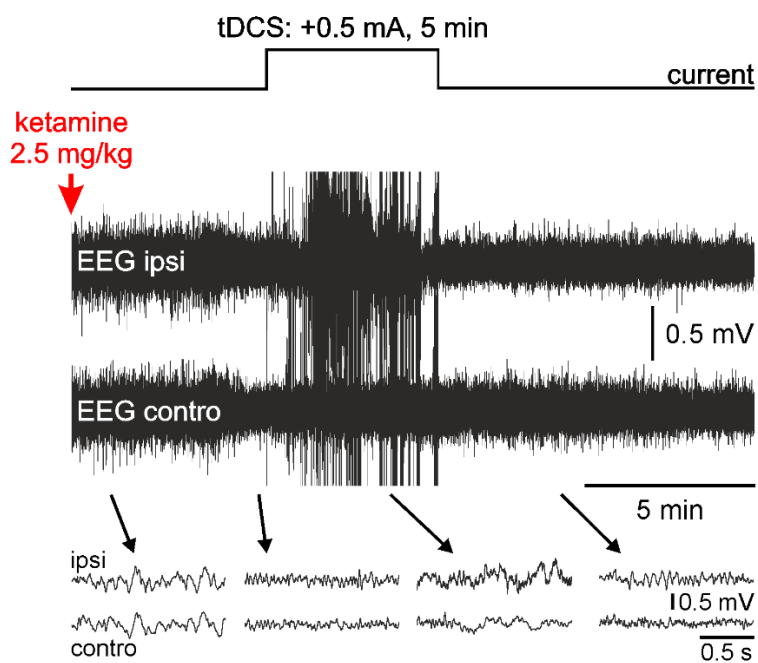

**Figure 4: Anodal (+0.5 mA) tDCS for 5 minutes modulates the EEG oscillation patterns.** In pentobarbital-sedated rats, the bilateral EEG exhibits a sleep-like pattern with slow oscillations, mainly slow waves in the delta frequency band (1-4 Hz), and brief (0.5 to 1.5 s) oscillations in the sigma band (10-17 Hz, spindle-like activities). Ketamine begins to transform EEG slow-waves into faster and less ample EEG waves 4 to 5 minutes after its systemic administration. The tDCS is applied 5 minutes after the ketamine administration. During the stimulation, the EEG ipsilateral to the stimulation electrodes is strongly altered with numerous artefacts.

**A**

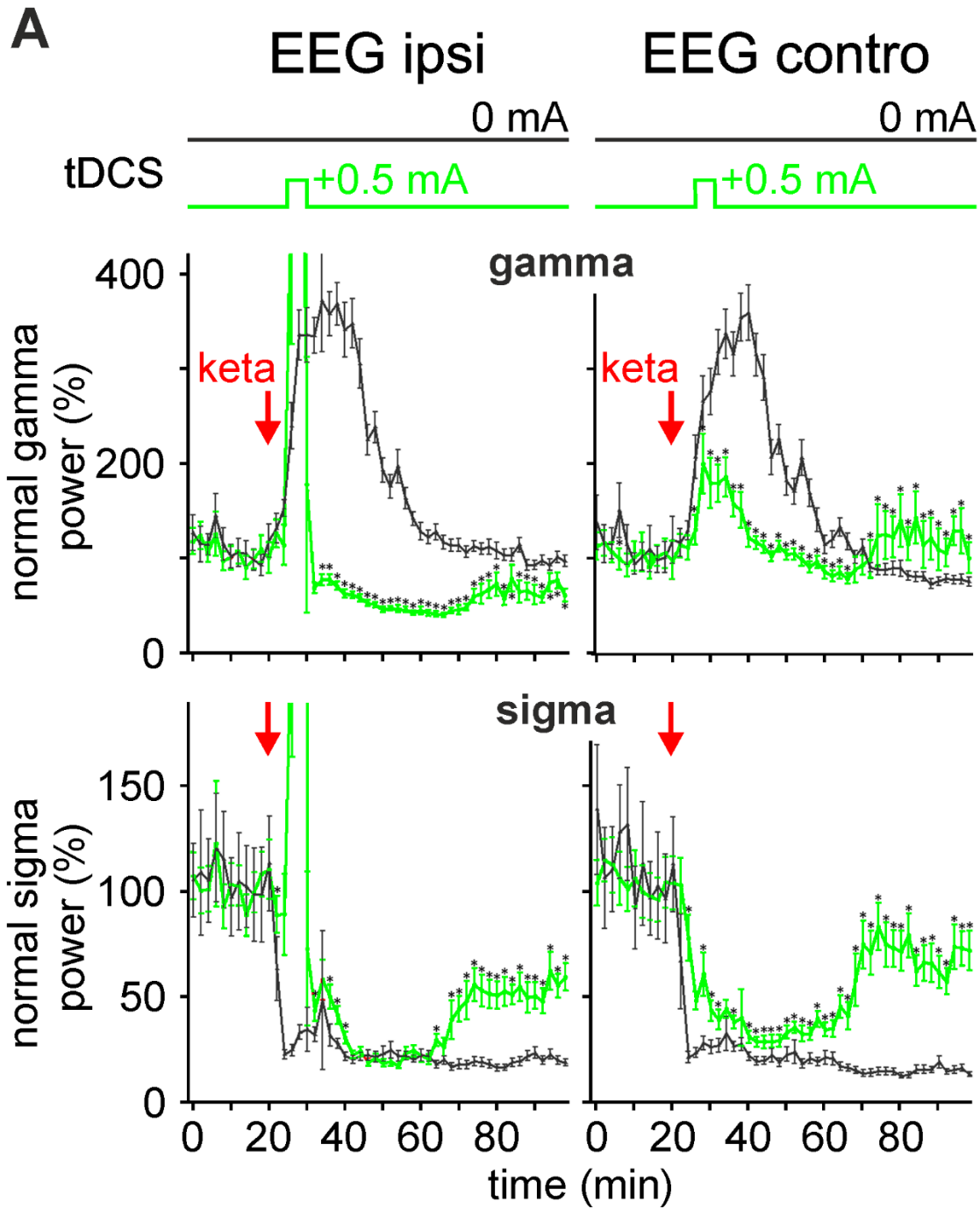

**Figure 5A: The bipolar anodal tDCS (+0.5 mA, 5 min) bilaterally reduces the ketamine peak effect.** In black ("sham control", 1 rat): Effect of ketamine (ketamine-tDCS 0 mA) on the normalized power of the gamma (top) and sigma (bottom) oscillations. In green (another rat): tDCS was applied 5 minutes after the subcutaneous injection of ketamine (2.5 mg/kg) and for 5 minutes (25 to 30 minutes). The values of the power of the oscillations are represented with a resolution of 2 minutes (each point is an average of 60 values, ± standard error of the average). The effect of the stimulation is compared to the ketamine-DCS (0 mA) control group. Student's test (*, $p < 0.05$).

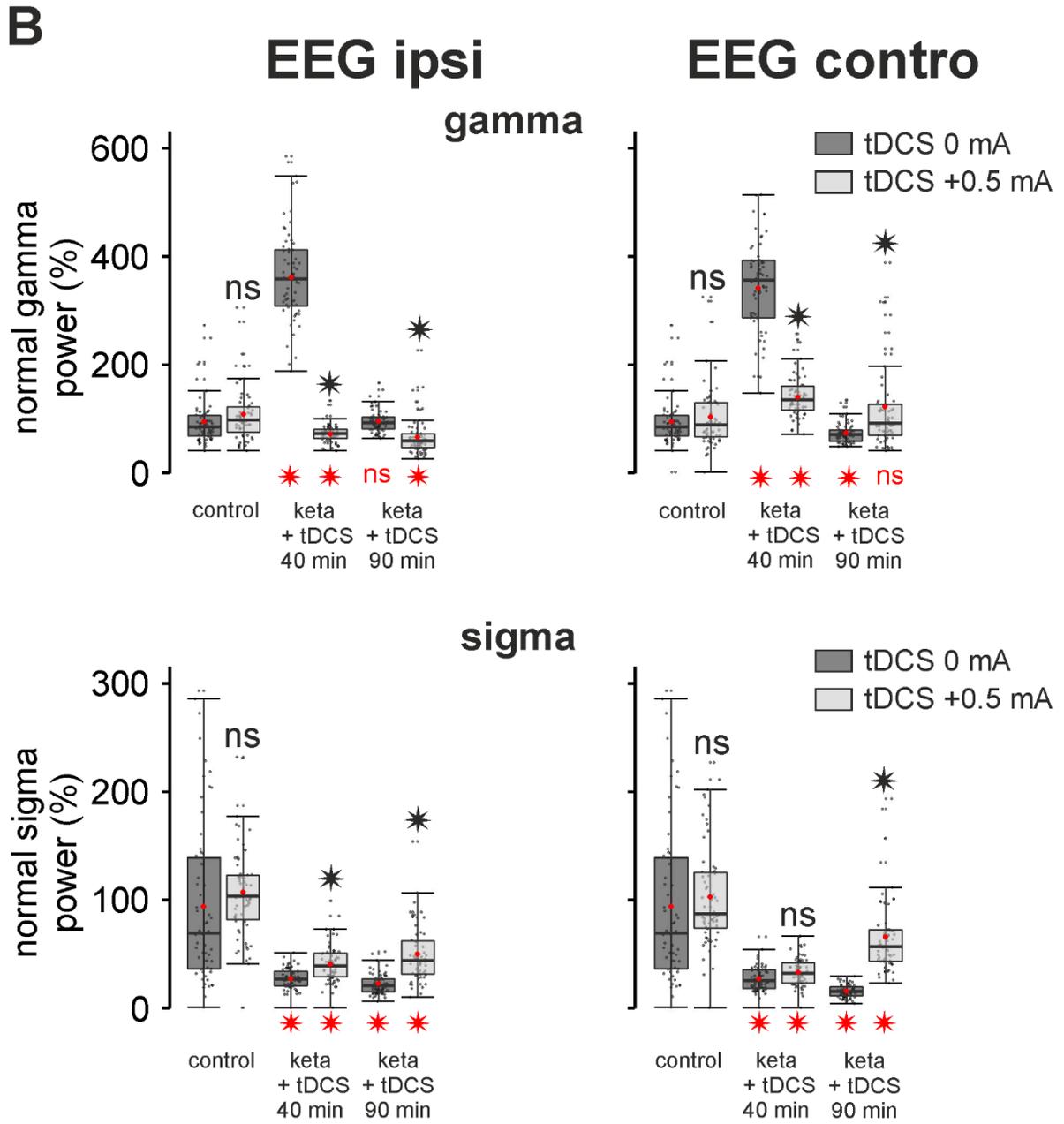

**Figure 5B: Boxplots from the data presented in the graphs of the figure 5A.** Red point: mean; thick black line: median; bottom and top of the box, first and third quartile, respectively; error bars: ± standard deviation. Two student's t-tests (ns, non-significant, *, p <0.05) were done. The one in black (above): comparison of ketamine-tDCS 0mA with ketamine-tDCS +0.5 mA. The red one (bottom): comparison of each box dataset to the saline condition (control, ~100%).

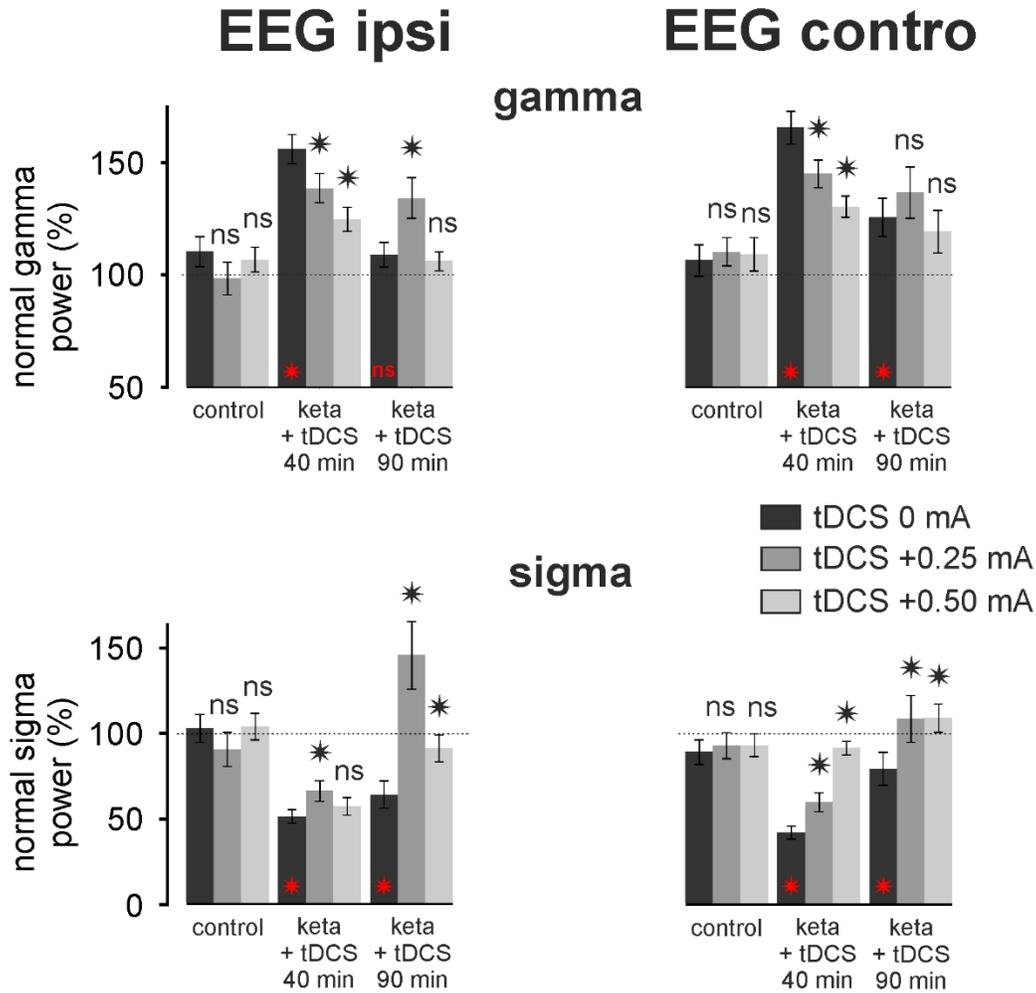

**Figure 6: Intensity effect of tDCS on the ketamine-induced oscillopathies.** The tDCS +0.25 or +0.50 mA was applied 8 minutes after the systemic administration of ketamine with 3 rats per condition (ketamine + tDCS 0, +0.25 or +0.50 mA). Each column represents the percentage of the normalized power of the gamma or sigma oscillations (average of 3 rats x 60 values over 2 minutes ± SEM) for each of the conditions: control (18-20 min), ketamine + tDCS 40 min (38-40 min) and at 90 min (88-90 min). T-test comparison, relative to the ketamine-tDCS 0 mA condition, of the ketamine-tDCS effect +0.25 or +0.50 mA (ns, not significant; * p <0.05). In red: T-test comparison, relative to the control condition (saline, 100%), of the ketamine effect (tDCS 0 mA) at 40 and 90 minutes after the ketamine administration.

### The minimum effective intensity.

Here, it is shown that exogenous currents had instantaneous influence on ongoing brain activities, which is in agreement with previous comprehensive studies[41, 42]. So, it was important to assess the minimum effective intensity in reducing or normalizing the ketamine-induced oscillopathies. For that, we compared the effects of the FP anodal tDCS +0.25 and +0.50 mA for 5 min with 3 rats per condition (Fig. 6). The tDCS was applied 8 minutes after the ketamine administration. A one-factor analysis of variance revealed an intensity effect in the bilateral EEG for both the ketamine-induced gamma hyperactivity (ipsilateral EEG: F = 23.48, p<0.001; controlateral EEG: F = 17.82, p<0.001) and the concomitant spindle hypoactivity (ipsilateral EEG: F = 42.94, p<0.001; controlateral EEG: F = 36.27, p<0.001). At 40 minutes, the tDCS was more effective with an intensity of +0.5 mA than +0.25 mA in reducing bilaterally the ketamine-induced gamma hyperactivity and in reducing controlaterally the concomitant spindle hypoactivity. A significant difference ($p<0.05$) between the +0.25 mA and +0.5 mA groups was observed. Because the tDCS was efficient in reducing the ketamine-induced spindle hypoactivity, we expected that the tDCS could similarly reduce the ketamine-induced delta hypoactivity. The figure 7 shows that the tDCS tended to diminish the delta hypoactivity. The efficacy was significant at 40 minutes for the intensity 0.5 mA in both sides and at 90 minutes for both intensities only in the ipsilateral EEG.

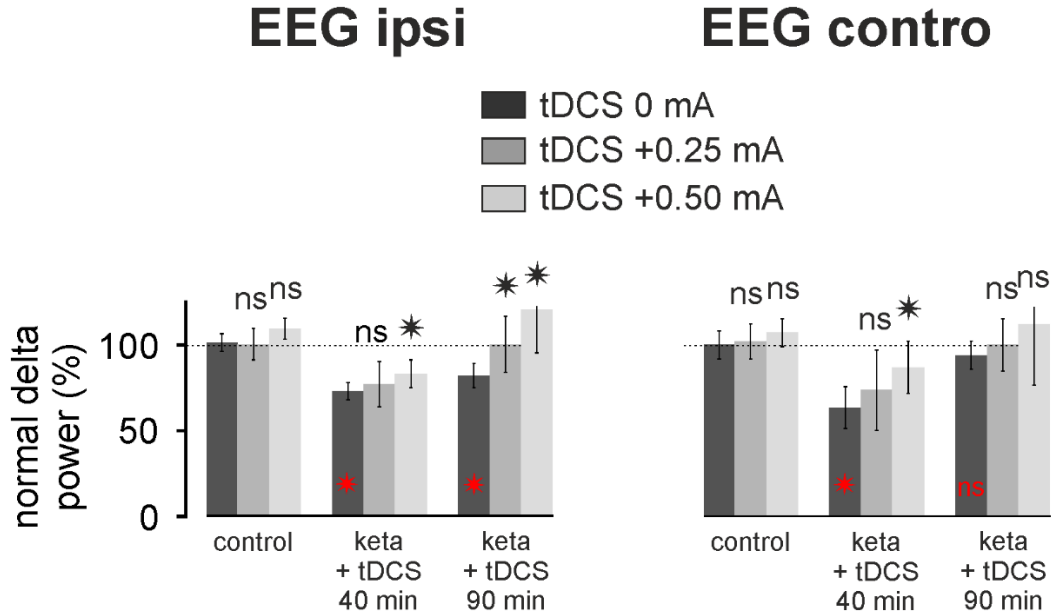

**Figure 7: The transcranial frontoparietal DCS tends to reduce the ketamine effects on delta oscillations.** The tDCS +0.25 or +0.50 mA was applied 8 minutes after the systemic administration of ketamine with 3 rats per condition (ketamine + tDCS 0, +0.25 or +0.50 mA). Each column represents the percentage of the normalized power of the delta oscillations (average of 3 rats x 60 values over 2 minutes ± SEM) for each of the conditions: control (18-20 min), ketamine + tDCS 40 min (38-40 min) and at 90 min (88-90 min). T-test comparison, relative to the ketamine-tDCS 0 mA condition, of the ketamine-tDCS effect +0.25 or +0.50 mA (ns, not significant; * p <0.05). In red: T-test comparison, relative to the control condition (saline, 100%), of the ketamine effect (tDCS 0 mA) at 40 and 90 minutes after the ketamine administration.

**DISCUSSION**

The present preclinical pilot investigation provides promising technical and neurophysiological essentials for developing a non-invasive therapeutic proof-of-concept against the transition to a psychotic state. A 5-min FP anodal tDCS with an intensity of less than +1 mA can, immediately and quickly, reduce, with intensity and duration effects, the ketamine-induced gamma hyperactivity and spindle hypoactivity. It tended also to reduce the associated delta hypoactivity. Technical, neurophysiological, neurochemical, and structural issues deserve discussions for the implementation of in-depth studies aiming at optimizing tDCS methods.

### Experimental conditions

Our experiments were carried out in rats lightly anesthetized with pentobarbital, which induces a slow-wave sleep with spindle-like activities by increasing the GABAergic neurotransmission[24, 43]. These experimental conditions, relatively stable over time (at least up to 8 hours) are "ideal" for determining and adjusting the multiple and various parameters of the electrical stimulation. Indeed, they made it possible to control, in a durable and relatively reliable way from one to another animal, a given parameter while recording the brain activities before, during, and after the stimulation. They also help us, over the course of the experiments, to optimize the stimulation parameters. Such a goal is almost impossible to swiftly achieve in the free-behaving animal, that is, to obtain a high success rate and reliable results in a reasonable time. Such a risky strategy would lead to a sacrifice of a large number of animals and would require years of research. The present pilot study may play a pivotal role in planning comprehensive studies at a reasonable cost.

Both the skin and the skull form a complex mechanical and bioelectric interface with the variability of skull conductivity and thickness, which can lead to inter-individual variability[39]. So, in our pilot investigation, we attempted to set up a simpler and more reliable preparation in an attempt to optimize, as precisely as possible, the multiple stimulation parameters (duration, intensity, polarity…). We securely placed our stimulation electrodes on saline-soaked sponges directly on the skull, which was slightly drilled at the electrode placement areas. Under these experimental conditions, a tDCS with an intensity of less than 1 mA was expected not to damage the structure and anatomical properties of intracortical neurons since the stimulation influence on the bilateral cortical EEG oscillations was reversible. This issue should, however, be histochemically validated.

### Technical considerations

In the present series of experiments, the unilateral FP tDCS was very useful to evaluate its local (ipsilateral EEG) and distant (controlateral EEG) effects. The latter effect very likely included a multi-synaptic influence mediated in great part by the callosal pathway. Regarding the recordings of the rat illustrated in the figure 5, it is clear that the tDCS +0.5 mA was immediately more effective in correcting the ketamine-induced gamma hyperactivity in the ipsilateral than the controlateral cortex. In contrast, the efficacy of the tDCS in reducing the ketamine-induced spindle hypoactivity was more evident in the controlateral than the ipsilateral cortical EEG. As a single 1-ms stimulation evoked more spindle-like activities in the controlateral than the ipsilateral cortex (see Fig. 1B1, B2), it would make sense to conclude that the efficacy of the tDCS on the controlateral spindle-like activities is secondary to the primary ipsilateral effects. Nevertheless, based on our previous data recorded under the same experimental conditions[24], it is questionable whether the late (~40 min after the ketamine administration) cortical activities were due either to a "true" stimulation effect (ketamine combined with tDCS +0.5 mA effects), or a spontaneous partial recovery (ketamine alone (tDCS 0 mA)). The present group data do not exclude an efficacy of the FP anodal tDCS paralleled with partial recovery. So, further investigation is necessary to clarify this point.

Furthermore, our results show, at the group level, an intensity effect (0, +0.25, and +0.50 mA, 3 rats per condition) of a 5-min FP anodal tDCS, indicating that the stimulation could partially or completely normalize the ketamine-induced oscillopathies, regarding at least the sigma- and broadband gamma-frequency oscillations. In humans, since the duration of the stimulation can be up to a few tens of minutes[34, 44, 45], it would be logical to seek a minimum effective intensity perhaps by increasing either the duration of the stimulation or the contact surface of the stimulating electrodes. Here, it was shown that the unilateral tDCS induced a functional imbalance between the two, ipsi- and controlateral cortices. Therefore, a bilateral stimulation may be expected to be a better alternative, perhaps with a lower intensity, which would correct this imbalance by influencing equally the neurophysiological activities on both sides.

The present study does not, nevertheless, offer a standardized tDCS method. Here, we assessed the effects of a unilateral, bipolar FP anodal tDCS on the acute ketamine-induced oscillopathies. With the bipolar format, further investigation is required to probe other options. For instance, as the frontal and parietal cortical areas are reciprocally connected, it would be logical to probe also a parieto-frontal anodal tDCS, the cathode at the parietal level and the anode at the frontal level. Also, knowing that anodal and cathodal tDCS can have similar effects on synaptic plasticity[46], it would be also wise to investigate the effects of FP and PF cathodal tDCS.

### Functional and mechanistic aspects

Under our experimental conditions, the unilateral, bipolar FP anodal tDCS exerted local and remote influences on ongoing cortically- and thalamically-generated activities. This is in agreement with the available relevant literature. Indeed, in humans, the DCS-induced electrical field spreads rapidly throughout the brain and modulates, locally and remotely, short- and long-range mono/multisynaptic cortical and cortical-subcortical systems[31-33, 47, 48]. Furthermore, it was demonstrated that anodal tDCS of the frontal cortex widespreadly increases cerebral blood flow in many cortical and subcortical structures[47], and can enhance sleepiness and sleep-related EEG oscillations[49]. However, the mechanisms underlying the immediate and downstream effects of regional tDCS remain to be elucidated. They may include immediate and in-cascade, short- and medium-term network, synaptic, cellular, and molecular processes, including multiple plasticity processes[30, 46, 50-52]. A computational study predicts that focal stimulation can trigger new functional large-scale neural connections[53]. The effects of the tDCS also depend on the state of the brain and neural systems[47, 54-59]. The thalamus might be implicated in the tDCS effects as high-frequency electrical stimulation of the FP thalamocortical pathway exerts effects that are opposite to the NMDA receptor antagonist ketamine, especially simultaneously on both the synaptic plasticity and the gamma power[22]. And experimental findings support the notion that electrical stimulation of the thalamus has pro-sensory/cognitive properties[60, 61].

Furthermore, from the EEG recordings of the parietal cortex, we can make testable predictions regarding the underlying cellular and synaptic activities of the thalamic neurons. Indeed, broadband gamma oscillations and spindles result from functional synaptic interactions between GABAergic and glutamatergic neurons. More specifically, gamma oscillations implicate such synaptic interactions in both the cortex and the thalamus[62, 63]; sleep-related spindles are generated principally in the thalamus with synaptic interactions between the thalamic relay and reticular neurons[24, 64]. So, when the EEG is synchronized, it predominantly displays delta oscillations and spindles, and the corresponding thalamic relay and reticular neurons mainly fire rhythmic high-frequency bursts of action potentials[64]. When the EEG is desynchronized, it predominantly exhibits faster and lower amplitude oscillations, including broadband gamma-frequency oscillations, and the corresponding thalamic relay and reticular neurons principally fire single action potentials in the tonic irregular mode. In the sedated rat, ketamine transiently reduces in power delta oscillations and spindles and increases in power broadband gamma- and higher-frequency oscillations by switching the firing of both relay and reticular neurons from the burst mode to the single action potential mode[24]. Therefore, we predict that, at least in the FP corticothalamic system of the ketamine-treated sedated rat, the FP anodal tDCS increased in power delta oscillations and spindles and decreased broadband gamma- and higher-frequency oscillations by switching the firing of the thalamic glutamatergic and GABAergic neurons from the single action potential mode to the burst mode. This prediction can be tested using a cell-to-network electrophysiological exploration (Fig. 8), which may help to understand some aspects of the mechanisms underlying the tDCS-induced change in the state of the concerned neural systems.

Previous studies support the notion that the tDCS can modulate the membrane activity of neurons[46], in particular through glutamatergic NMDA receptors and GABAergic receptors[65, 66]. The anodal stimulation would depolarize the membrane potential and increase neuronal excitability, in particular by reducing intracortical inhibition and increasing paired-pulse excitability[45]. On the other hand, cathodal stimulation, would hyperpolarize the membrane potential and decrease neuronal excitability through its inhibitory action on glutamatergic neurons[65-67].

The blockade of NMDA receptors by ketamine prevents the expression of the long-term potentiation in rodents[22]. The tDCS can modulate the synaptic plasticity with effects that depend on the spatial and temporal properties of synapses[46, 68]. We can hypothesize that since tDCS can reduce the abnormal oscillations induced by ketamine, it will also be able to durably reduce the ketamine-induced decrease in LTP, in turn modulating LTP or LTD through the glutamatergic and GABAergic receptors. Further investigation is necessary to decipher the mechanisms underlying the impact of the FP anodal tDCS on the ketamine-induced oscillopathies.

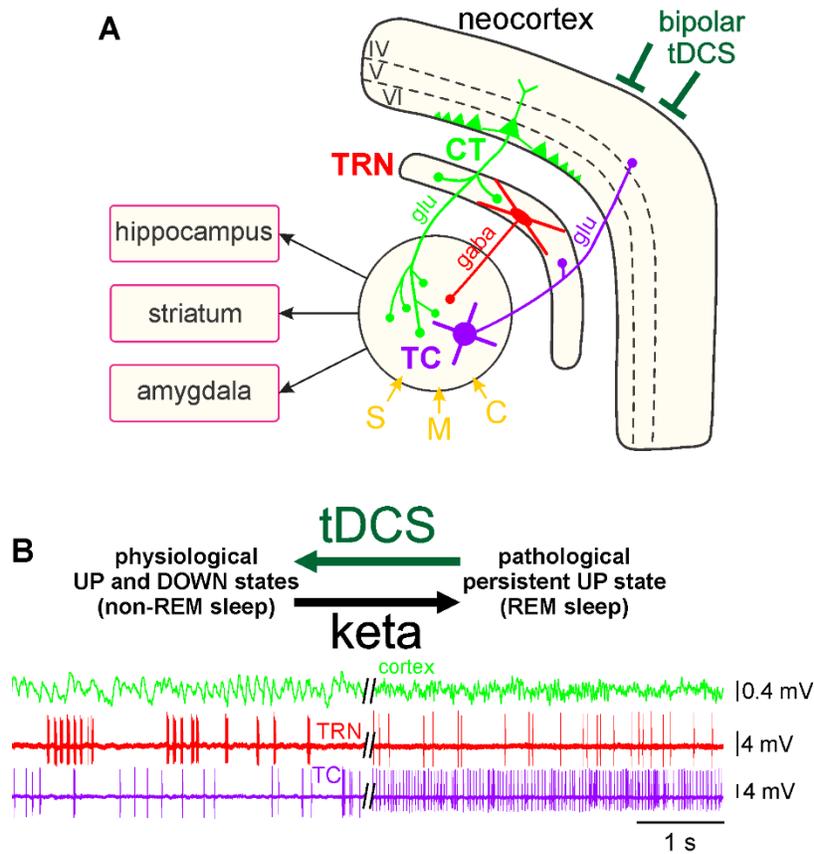

**Figure 8: Theoretical prediction of the cell-to-network effects of a frontoparietal anodal tDCS in the corticothalamic system. (A)** Simplified drawing of the hodology of the 3-neuron CT-TRN-TC circuit. The layer 6 corticothalamic (CT) and thalamocortical (TC) neurons are glutamatergic while the thalamic reticular nucleus (TRN) neuron is GABAergic. Both TC and CT axons innervate the TRN. This system receives sensory (S), motor (M) and cognitive/associative (C) inputs. It is important to specify that the layer VI CT neurons outnumber by a factor of about 10 the TC neurons. **(B, left)** Physiological UP and DOWN states: During the non-REM sleep, the TC system displays principally a synchronized state, characterized by the occurrence of delta oscillations and spindles; the TRN cell exhibits mainly rhythmic (at the delta-, theta- and spindle-frequency bands) high-frequency bursts of action potentials. The synchronized state includes two sub-states, UP and DOWN, which are usually associated with active and quiescent cellular firings, respectively. **(B, right)** Pathological persistent UP state: This ketamine-induced persistent UP state is assumed to be an abnormal REM sleep. After a single systemic administration of a subanesthetizing low-dose of ketamine, the TC system displays a more desynchronized state (peak effect at about +15-20 minutes) characterized by the prominent occurrence of lower voltage and faster activities (>16 Hz), which include beta-, gamma- and higher-frequency oscillations. Under the ketamine condition, both the TC and the TRN neurons exhibit a persistent irregular and tonic firing containing more single APs than high-frequency bursts of APs. The bipolar anodal tDCS is expected to reduce, even to normalize, the ketamine-induced oscillopathies. Adapted from Mahdavi et al., Schizophr Res, 2020.

**Outlook**

The present, conceptually- and data-driven pilot study brings key findings that may help, through comprehensive studies, the standardization of tDCS methods in animal models of psychosis transition and individuals having a high-risk state for psychosis.

At the preclinical level, after the primary phase of scientific and technological innovations, it would be rational to investigate the unilateral or bilateral tDCS in the ongoing brain activities of free-behaving rodents under physiological and pathological (e.g, under ketamine influence) conditions. The atypical antipsychotic clozapine is efficient in reducing, at least in the rodent, the ketamine- or MK-801-induced oscillopathies[24, 69-71]. Therefore, the acute ketamine model seems appropriate to assess the potential preventive and/or curative effects of tDCS. In the second phase of research and development, it would be logical to explore the tDCS in more realistic models of psychotic transition, for instance of genetic-neurodevelopmental types. This would certainly help to appreciate whether the tDCS could correct or normalize not only the brain oscillopathies but also the associated psychosis-relevant behavior and cognitive impairment. Indeed and interestingly, tDCS during adolescence, before psychosis-relevant behavioral abnormalities, prevents the development of positive symptoms in the rodent maternal immune stimulation model of schizophrenia[72].

The timing of the stimulation application is an important parameter that may condition the efficacy of the tDCS. For instance, in the acute ketamine model of psychosis transition, it would be essential to know a threshold value of the ongoing brain or large-scale network oscillopathies, from which a closed-loop device could trigger, uniquely or repetitively, the stimulator. The threshold value could be, for instance, the amplitude, the power, and/or the pattern of the aberrant oscillatory activities. Thereby, closed-loop neurostimulation would provide therapeutic stimulation only when necessary, a promising way for oscillotherapeutics[73] in a frame of personalized medicine. Early studies have shown that diverse conditions such as Parkinson's disease[74], chronic pain[75], and intractable temporal lobe of epilepsy[76] can benefit from a therapeutic closed-loop stimulation.

## ACKNOWLEDGEMENTS

The present work was supported by INSERM, the French National Institute of Health and Medical Research (Institut National de la Santé et de la Recherche Médicale, 2013-), l'Université de Strasbourg, Unistra (2013-), and Neurex. CL was a graduate student from The University of Strasbourg (October 2019-June 2020). The authors thank Laura Winkler from the Joint Master in Neuroscience, The University of Strasbourg, for her contribution in data analysis, and Damaris Cornec for her technical assistance throughout the experiments.

## DISCLOSURES

The authors have approved the final version of the article. And they report no competing biomedical financial interests or potential conflicts of interest.

## AUTHORS' CONTRIBUTION

CL, DP: Design, data acquisition & analysis, and writing.

# REFERENCES

[1] Patel KR, Cherian J, Gohil K, Atkinson D. Schizophrenia: overview and treatment options. P T 2014;39(9):638-45.

[2] Uno Y, Coyle JT. Glutamate hypothesis in schizophrenia. Psychiatry Clin Neurosci 2019;73(5):204-15.

[3] Jahshan C, Heaton RK, Golshan S, Cadenhead KS. Course of neurocognitive deficits in the prodrome and first episode of schizophrenia. Neuropsychology 2010;24(1):109-20.

[4] Kambeitz J, Kambeitz-Ilankovic L, Cabral C, Dwyer DB, Calhoun VD, van den Heuvel MP, et al. Aberrant Functional Whole-Brain Network Architecture in Patients With Schizophrenia: A Meta-analysis. Schizophr Bull 2016;42 Suppl 1:S13-21.

[5] Lord LD, Allen P, Expert P, Howes O, Broome M, Lambiotte R, et al. Functional brain networks before the onset of psychosis: A prospective fMRI study with graph theoretical analysis. Neuroimage Clin 2012;1(1):91-8.

[6] Uhlhaas PJ, Singer W. Neuronal dynamics and neuropsychiatric disorders: toward a translational paradigm for dysfunctional large-scale networks. Neuron 2012;75(6):963-80.

[7] Herrmann CS, Demiralp T. Human EEG gamma oscillations in neuropsychiatric disorders. Clin Neurophysiol 2005;116(12):2719-33.

[8] Ramyead A, Kometer M, Studerus E, Koranyi S, Ittig S, Gschwandtner U, et al. Aberrant Current Source-Density and Lagged Phase Synchronization of Neural Oscillations as Markers for Emerging Psychosis. Schizophr Bull 2015;41(4):919-29.

[9] Castelnovo A, Graziano B, Ferrarelli F, D'Agostino A. Sleep spindles and slow waves in schizophrenia and related disorders: main findings, challenges and future perspectives. Eur J Neurosci 2018;48(8):2738-58.

[10] Kaskie RE, Ferrarelli F. Investigating the neurobiology of schizophrenia and other major psychiatric disorders with Transcranial Magnetic Stimulation. Schizophr Res 2018;192:30-8.

[11] Ferrarelli F, Peterson MJ, Sarasso S, Riedner BA, Murphy MJ, Benca RM, et al. Thalamic dysfunction in schizophrenia suggested by whole-night deficits in slow and fast spindles. Am J Psychiatry 2010;167(11):1339-48.

[12] Ferrarelli F, Huber R, Peterson MJ, Massimini M, Murphy M, Riedner BA, et al. Reduced sleep spindle activity in schizophrenia patients. Am J Psychiatry 2007;164(3):483-92.

[13] Manoach DS, Pan JQ, Purcell SM, Stickgold R. Reduced Sleep Spindles in Schizophrenia: A Treatable Endophenotype That Links Risk Genes to Impaired Cognition? Biol Psychiatry 2016;80(8):599-608.

[14] Manoach DS, Demanuele C, Wamsley EJ, Vangel M, Montrose DM, Miewald J, et al. Sleep spindle deficits in antipsychotic-naive early course schizophrenia and in non-psychotic first-degree relatives. Front Hum Neurosci 2014;8:762.

[15] Fusar-Poli P, Crossley N, Woolley J, Carletti F, Perez-Iglesias R, Broome M, et al. White matter alterations related to P300 abnormalities in individuals at high risk for psychosis: an MRI-EEG study. J Psychiatry Neurosci 2011;36(4):239-48.

[16] Stone JM, Day F, Tsagaraki H, Valli I, McLean MA, Lythgoe DJ, et al. Glutamate dysfunction in people with prodromal symptoms of psychosis: relationship to gray matter volume. Biol Psychiatry 2009;66(6):533-9.

[17] Howes O, McCutcheon R, Stone J. Glutamate and dopamine in schizophrenia: an update for the 21st century. J Psychopharmacol 2015;29(2):97-115.

[18] Kocsis B. Differential role of NR2A and NR2B subunits in N-methyl-D-aspartate receptor antagonist-induced aberrant cortical gamma oscillations. Biol Psychiatry 2012;71(11):987-95.

[19] Pinault D. N-methyl d-aspartate receptor antagonists ketamine and MK-801 induce wake-related aberrant gamma oscillations in the rat neocortex. Biol Psychiatry 2008;63(8):730-5.

[20] Hakami T, Jones NC, Tolmacheva EA, Gaudias J, Chaumont J, Salzberg M, et al. NMDA receptor hypofunction leads to generalized and persistent aberrant gamma oscillations independent of hyperlocomotion and the state of consciousness. PLoS One 2009;4(8):e6755.

[21] Rivolta D, Heidegger T, Scheller B, Sauer A, Schaum M, Birkner K, et al. Ketamine Dysregulates the Amplitude and Connectivity of High-Frequency Oscillations in Cortical-Subcortical Networks in Humans: Evidence From Resting-State Magnetoencephalography-Recordings. Schizophr Bull 2015;41(5):1105-14.

[22] Kulikova SP, Tolmacheva EA, Anderson P, Gaudias J, Adams BE, Zheng T, et al. Opposite effects of ketamine and deep brain stimulation on rat thalamocortical information processing. Eur J Neurosci 2012;36(10):3407-19.

[23] Anderson PM, Jones NC, O'Brien TJ, Pinault D. The N-Methyl d-Aspartate Glutamate Receptor Antagonist Ketamine Disrupts the Functional State of the Corticothalamic Pathway. Cereb Cortex 2017;27(6):3172-85.

[24] Mahdavi A, Qin Y, Aubry AS, Cornec D, Kulikova S, Pinault D. A single psychotomimetic dose of ketamine decreases thalamocortical spindles and delta oscillations in the sedated rat. Schizophr Res 2020. https://doi.org/10.1016/j.schres.2020.04.029.

[25] Kane JM, Honigfeld G, Singer J, Meltzer H. Clozapine in treatment-resistant schizophrenics. Psychopharmacol Bull 1988;24(1):62-7.

[26] Okhuijsen-Pfeifer C, Huijsman EAH, Hasan A, Sommer IEC, Leucht S, Kahn RS, et al. Clozapine as a first- or second-line treatment in schizophrenia: a systematic review and meta-analysis. Acta Psychiatr Scand 2018;138(4):281-8.

[27] Bennabi D, Pedron S, Haffen E, Monnin J, Peterschmitt Y, Van Waes V. Transcranial direct current stimulation for memory enhancement: from clinical research to animal models. Front Syst Neurosci 2014;8:159.

[28] Bikson M, Grossman P, Thomas C, Zannou AL, Jiang J, Adnan T, et al. Safety of Transcranial Direct Current Stimulation: Evidence Based Update 2016. Brain Stimul 2016;9(5):641-61.

[29] Coffman BA, Clark VP, Parasuraman R. Battery powered thought: enhancement of attention, learning, and memory in healthy adults using transcranial direct current stimulation. Neuroimage 2014;85 Pt 3:895-908.

[30] Nitsche MA, Cohen LG, Wassermann EM, Priori A, Lang N, Antal A, et al. Transcranial direct current stimulation: State of the art 2008. Brain Stimul 2008;1(3):206-23.

[31] Chhatbar PY, Kautz SA, Takacs I, Rowland NC, Revuelta GJ, George MS, et al. Evidence of transcranial direct current stimulation-generated electric fields at subthalamic level in human brain in vivo. Brain Stimul 2018;11(4):727-33.

[32] Dalong G, Jiyuan L, Ying Z, Lei Z, Yanhong H, Yongcong S. Transcranial direct current stimulation reconstructs diminished thalamocortical connectivity during prolonged resting wakefulness: a resting-state fMRI pilot study. Brain Imaging Behav 2020;14(1):278-88.

[33] Polania R, Paulus W, Nitsche MA. Modulating cortico-striatal and thalamo-cortical functional connectivity with transcranial direct current stimulation. Hum Brain Mapp 2012;33(10):2499-508.

[34] Mondino M, Jardri R, Suaud-Chagny MF, Saoud M, Poulet E, Brunelin J. Effects of Fronto-Temporal Transcranial Direct Current Stimulation on Auditory Verbal Hallucinations and Resting-State Functional Connectivity of the Left Temporo-Parietal Junction in Patients With Schizophrenia. Schizophr Bull 2016;42(2):318-26.

[35] Valiengo L, Goerigk S, Gordon PC, Padberg F, Serpa MH, Koebe S, et al. Efficacy and Safety of Transcranial Direct Current Stimulation for Treating Negative Symptoms in Schizophrenia: A Randomized Clinical Trial. JAMA Psychiatry 2019.

[36] Hoy KE, Bailey NW, Arnold SL, Fitzgerald PB. The effect of transcranial Direct Current Stimulation on gamma activity and working memory in schizophrenia. Psychiatry Res 2015;228(2):191-6.

[37] Rabanea-Souza T, Cirigola SMC, Noto C, Gomes JS, Azevedo CC, Gadelha A, et al. Evaluation of the efficacy of transcranial direct current stimulation in the treatment of cognitive symptomatology in the early stages of psychosis: study protocol for a double-blind randomized controlled trial. Trials 2019;20(1):199.

[38] Kovac S, Kahane P, Diehl B. Seizures induced by direct electrical cortical stimulation--Mechanisms and clinical considerations. Clin Neurophysiol 2016;127(1):31-9.

[39] Antonakakis M, Schrader S, Aydin U, Khan A, Gross J, Zervakis M, et al. Inter-Subject Variability of Skull Conductivity and Thickness in Calibrated Realistic Head Models. Neuroimage 2020;223:117353.

[40] Pinault D, Slezia A, Acsady L. Corticothalamic 5-9 Hz oscillations are more pro-epileptogenic than sleep spindles in rats. J Physiol 2006;574(Pt 1):209-27.

[41] Fruhauf AMA, Politti F, Dal Corso S, Costa GC, Teodosio ADC, Silva SM, et al. Immediate effect of transcranial direct current stimulation combined with functional electrical stimulation on activity of the tibialis anterior muscle and balance of individuals with hemiparesis stemming from a stroke. J Phys Ther Sci 2017;29(12):2138-46.

[42] Liu A, Voroslakos M, Kronberg G, Henin S, Krause MR, Huang Y, et al. Immediate neurophysiological effects of transcranial electrical stimulation. Nat Commun 2018;9(1):5092.

[43] Maldifassi MC, Baur R, Sigel E. Functional sites involved in modulation of the GABAA receptor channel by the intravenous anesthetics propofol, etomidate and pentobarbital. Neuropharmacology 2016;105:207-14.

[44] Palm U, Keeser D, Hasan A, Kupka MJ, Blautzik J, Sarubin N, et al. Prefrontal Transcranial Direct Current Stimulation for Treatment of Schizophrenia With Predominant Negative Symptoms: A Double-Blind, Sham-Controlled Proof-of-Concept Study. Schizophr Bull 2016;42(5):1253-61.

[45] Woods AJ, Antal A, Bikson M, Boggio PS, Brunoni AR, Celnik P, et al. A technical guide to tDCS, and related non-invasive brain stimulation tools. Clin Neurophysiol 2016;127(2):1031-48.

[46] Kronberg G, Bridi M, Abel T, Bikson M, Parra LC. Direct Current Stimulation Modulates LTP and LTD: Activity Dependence and Dendritic Effects. Brain Stimul 2017;10(1):51-8.

[47] Lang N, Siebner HR, Ward NS, Lee L, Nitsche MA, Paulus W, et al. How does transcranial DC stimulation of the primary motor cortex alter regional neuronal activity in the human brain? Eur J Neurosci 2005;22(2):495-504.

[48] Nitsche MA, Seeber A, Frommann K, Klein CC, Rochford C, Nitsche MS, et al. Modulating parameters of excitability during and after transcranial direct current stimulation of the human motor cortex. J Physiol 2005;568(Pt 1):291-303.

[49] D'Atri A, De Simoni E, Gorgoni M, Ferrara M, Ferlazzo F, Rossini PM, et al. Electrical stimulation of the frontal cortex enhances slow-frequency EEG activity and sleepiness. Neuroscience 2016.

[50] Krause B, Marquez-Ruiz J, Cohen Kadosh R. The effect of transcranial direct current stimulation: a role for cortical excitation/inhibition balance? Front Hum Neurosci 2013;7:602.

[51] Medeiros LF, de Souza IC, Vidor LP, de Souza A, Deitos A, Volz MS, et al. Neurobiological effects of transcranial direct current stimulation: a review. Front Psychiatry 2012;3:110.

[52] Pinault D. A Neurophysiological Perspective on a Preventive Treatment against Schizophrenia Using Transcranial Electric Stimulation of the Corticothalamic Pathway. Brain Sci 2017;7(4).

[53] Lu H, Gallinaro JV, Rotter S. Network remodeling induced by transcranial brain stimulation: A computational model of tDCS-triggered cell assembly formation. Netw Neurosci 2019;3(4):924-43.

[54] Benwell CS, Learmonth G, Miniussi C, Harvey M, Thut G. Non-linear effects of transcranial direct current stimulation as a function of individual baseline performance: Evidence from biparietal tDCS influence on lateralized attention bias. Cortex 2015;69:152-65.

[55] Fertonani A, Miniussi C. Transcranial Electrical Stimulation: What We Know and Do Not Know About Mechanisms. Neuroscientist 2016.

[56] Hill AT, Fitzgerald PB, Hoy KE. Effects of Anodal Transcranial Direct Current Stimulation on Working Memory: A Systematic Review and Meta-Analysis of Findings From Healthy and Neuropsychiatric Populations. Brain Stimul 2015.

[57] Marshall L, Kirov R, Brade J, Molle M, Born J. Transcranial electrical currents to probe EEG brain rhythms and memory consolidation during sleep in humans. PLoS One 2011;6(2):e16905.

[58] Nitsche MA, Fricke K, Henschke U, Schlitterlau A, Liebetanz D, Lang N, et al. Pharmacological modulation of cortical excitability shifts induced by transcranial direct current stimulation in humans. J Physiol 2003;553(Pt 1):293-301.

[59] Stagg CJ, Jayaram G, Pastor D, Kincses ZT, Matthews PM, Johansen-Berg H. Polarity and timing-dependent effects of transcranial direct current stimulation in explicit motor learning. Neuropsychologia 2011;49(5):800-4.

[60] Mair RG, Hembrook JR. Memory enhancement with event-related stimulation of the rostral intralaminar thalamic nuclei. J Neurosci 2008;28(52):14293-300.

[61] Shirvalkar P, Seth M, Schiff ND, Herrera DG. Cognitive enhancement with central thalamic electrical stimulation. Proc Natl Acad Sci U S A 2006;103(45):17007-12.

[62] Minlebaev M, Colonnese M, Tsintsadze T, Sirota A, Khazipov R. Early gamma oscillations synchronize developing thalamus and cortex. Science 2011;334(6053):226-9.

[63] Pinault D, Deschênes M. Voltage-dependent 40-Hz oscillations in rat reticular thalamic neurons in vivo. Neuroscience 1992;51(2):245-58.

[64] Steriade M, McCormick DA, Sejnowski TJ. Thalamocortical oscillations in the sleeping and aroused brain. Science 1993;262(5134):679-85.

[65] Liebetanz D, Nitsche MA, Tergau F, Paulus W. Pharmacological approach to the mechanisms of transcranial DC-stimulation-induced after-effects of human motor cortex excitability. Brain 2002;125(Pt 10):2238-47.

[66] Stagg CJ, Best JG, Stephenson MC, O'Shea J, Wylezinska M, Kincses ZT, et al. Polarity-sensitive modulation of cortical neurotransmitters by transcranial stimulation. J Neurosci 2009;29(16):5202-6.

[67] Nitsche MA, Paulus W. Excitability changes induced in the human motor cortex by weak transcranial direct current stimulation. J Physiol 2000;527 Pt 3:633-9.

[68] Monte-Silva K, Kuo MF, Hessenthaler S, Fresnoza S, Liebetanz D, Paulus W, et al. Induction of late LTP-like plasticity in the human motor cortex by repeated non-invasive brain stimulation. Brain Stimul 2013;6(3):424-32.

[69] Hunt MJ, Olszewski M, Piasecka J, Whittington MA, Kasicki S. Effects of NMDA receptor antagonists and antipsychotics on high frequency oscillations recorded in the nucleus accumbens of freely moving mice. Psychopharmacology (Berl) 2015;232(24):4535-35.

[70] Jones NC, Reddy M, Anderson P, Salzberg MR, O'Brien TJ, Pinault D. Acute administration of typical and atypical antipsychotics reduces EEG gamma power, but only the preclinical compound LY379268 reduces the ketamine-induced rise in gamma power. Int J Neuropsychopharmacol 2012;15(5):657-68.

[71] Anderson PM, Pinault D, O'Brien TJ, Jones NC. Chronic administration of antipsychotics attenuates ongoing and ketamine-induced increases in cortical gamma oscillations. Int J Neuropsychopharmacol 2014;17(11):1895-904.

[72] Hadar R, Winter R, Edemann-Callesen H, Wieske F, Habelt B, Khadka N, et al. Prevention of schizophrenia deficits via non-invasive adolescent frontal cortex stimulation in rats. Mol Psychiatry 2020;25(4):896-905.

[73] Takeuchi Y, Berenyi A. Oscillotherapeutics - Time-targeted interventions in epilepsy and beyond. Neurosci Res 2020;152:87-107.

[74] Velisar A, Syrkin-Nikolau J, Blumenfeld Z, Trager MH, Afzal MF, Prabhakar V, et al. Dual threshold neural closed loop deep brain stimulation in Parkinson disease patients. Brain Stimul 2019;12(4):868-76.

[75] Russo M, Cousins MJ, Brooker C, Taylor N, Boesel T, Sullivan R, et al. Effective Relief of Pain and Associated Symptoms With Closed-Loop Spinal Cord Stimulation System: Preliminary Results of the Avalon Study. Neuromodulation 2018;21(1):38-47.

[76] Geller EB, Skarpaas TL, Gross RE, Goodman RR, Barkley GL, Bazil CW, et al. Brain-responsive neurostimulation in patients with medically intractable mesial temporal lobe epilepsy. Epilepsia 2017;58(6):994-1004.